\documentclass[preprint]{elsarticle}
\usepackage[version=4]{mhchem}
\usepackage[utf8]{inputenc}
\usepackage{times}
\usepackage{graphicx}
\usepackage{color}
\usepackage{amsmath}
\usepackage{amssymb}
\usepackage{bm}

\begin{document}

\title{Gas Diffusion in Cement Pastes: An Analysis using a Fluctuating Diffusivity Model}

\author[nu]{Fumiaki Nakai}
\ead{nakai.fumiaki.c7@s.mail.nagoya-u.ac.jp}
\author[nu]{Takato Ishida \corref{cor1}}
\ead{ishida@mp.pse.nagoya-u.ac.jp}
\address[nu]{Department of Materials Physics, Graduate School of Engineering, Nagoya University, Furo-cho, Chikusa, Nagoya 464-8603, Japan}
\cortext[cor1]{Corresponding author}

\begin{abstract}
This study proposes applying the concept of fluctuating diffusivity (FD) to the diffusion of gas molecules in cementitious materials, mainly through a two-state fluctuating diffusivity (2SFD) model.
The 2SFD model was utilized to investigate oxygen diffusion in cement pastes.
This analysis provides a reasonable description of the oxygen diffusion coefficient in cement pastes and suggests the presence of non-Gaussian diffusion, which is attributed to the heterogeneous microstructure.
The presence of non-Gaussianity in the probability density of the displacement of the molecule, characterized by heavier tails than those of the Gaussian distribution, may significantly impact the durability of concrete structures.
\end{abstract}


\maketitle

\section{Introduction}
Since the invention of Portland cement by Joseph Aspdin in 1824, cementitious materials have been widely utilized in infrastructure construction.
In recent decades, there has been a growing emphasis on assessing the long-term performance of reinforced concrete structures with a focus on reducing carbon emissions and preserving resources.
The penetration of aggressive, lightweight molecules can compromise the durability of concrete structures (causing chemical degradation \cite{Glasser2008} such as carbonation \cite{Chang2006, Bary2004, Morandeau2014}, corrosion \cite{Angst2009,Ma2015}, sulfate attack \cite{Neville2004}, and calcium leaching \cite{Carde1996,Carde1997}), making the study of transport phenomena in cementitious materials a vital subject in the field of cement and concrete research.
It is evident that cementitious materials are inherently porous, possessing pores of various scales. Diffusion, the primary mode of mass transport, has been determined by devising effective diffusion coefficients that appropriately reflect the characteristics of the pore network structure (tortuosity, connectivity, constrictivity, and formation factor \cite{Promentilla2009,Patel2016,Hatanaka2017,Yio2019, Yang2020, Gao2021}) and utilizing them to solve the diffusion equation.
The probabilistic displacement distribution is Gaussian when the conventional diffusion equation is resolved \cite{Zwanzig2001, Nelson2020}.
However, recent research in theoretical physics has highlighted cases in which the displacement distribution deviates from a Gaussian distribution depending on the spatiotemporal scale of interest.
Such non-Gaussianity may significantly impact the long-term reliability assessment of reinforced-concrete structures. We effectively formulated the concept in a form that applies to diffusion in cementitious materials.

The microstructure of cementitious materials inherently exhibits a heterogeneous composition, which can result in the non-Gaussian diffusion of gases.
To effectively describe the diffusion in heterogeneous materials, the concept of fluctuating diffusivity (FD) \cite{metzler2020superstatistics, chechkin2017brownian, chubynsky2014diffusing, uneyama2019relaxation, uneyama2015fluctuation, miyaguchi2016langevin, miyaguchi2017elucidating, jain2017diffusing} has been demonstrated to be helpful, as evidenced by studies on glass-forming liquids \cite{rusciano2022fickian}, colloidal suspensions \cite{pastore2021rapid, kim2013simulation}, and biological systems \cite{jeon2016protein, wang2009anomalous}.
The diffusion of a free molecule with fluctuating diffusivity is described by the equation:

\begin{equation}
\frac{\partial G(\bm{r}; t)}{\partial t}
=D(t)\nabla^2 G(\bm{r};t)
\label{eq:diffusion_equation_with_fd}
\end{equation}

where $t$ denotes time, $\bm{r}$ represents the displacement vector of the particle, $G(\bm{r}; t)$ is the probability density of $\bm{r}$ at time $t$, and $D(t)$ represents the fluctuating diffusivity and is subject to a stochastic process. By providing a simple and physically reasonable rule for $D(t)$, it is possible to analyze the dynamics of diffusing particles theoretically.
Fluctuating diffusivity is based on the idea that the diffusion environment experienced by the particle changes over time, either as a result of a temporal alteration in the environment or owing to the migration of particles to a distinct milieu.
Upon initial inspection, the fluctuating diffusivity approach, expressed in Eq. ~\eqref{eq:diffusion_equation_with_fd}, may appear  similar to time-dependent diffusivity models that consider the long-term effects of changing diffusion media, such as prolonged hydration reactions and accumulated damages \cite{hristov2022diffusion, Yu2016, Xu2020}. However, it is essential to note that these two approaches are fundamentally distinct in terms of their concepts and underlying motivations.
The fluctuating diffusivity approach posits that the diffusion coefficient changes stochastically over time, thereby reflecting the temporal and spatial heterogeneity of the matrix. By contrast, the time-dependent diffusion coefficient varies deterministically, reflecting the time evolution of the internal microstructures caused by long-term effects.
In this study, the latter approach is referred to as deterministic drifting diffusivity (DDD) and is distinguished from fluctuating diffusivity by the deterministic variation of the diffusion coefficient.
Undeniably, the extensive research conducted on DDD has dramatically enhanced our understanding of the transport phenomena in cementitious materials and continues to be applied effectively in current studies.
It is important to note that fluctuating diffusivity does not aim to replace or update DDD; instead, it takes a distinct physical perspective.
The target timescale is significantly different between the fluctuating diffusivity and DDD approaches.
Typically, FD analyzes the particle diffusion on a timescale where the particle diffuses over the characteristic length of the heterogeneous environment. In contrast, the DDD approach focuses on the timescale where the state of the diffusion medium changes over a prolonged period.
It is important to note that some studies have employed the DDD approach \cite{hristov2022diffusion}, which does not consider temporal and spatial fluctuations and may need to be revised to describe diffusion in heterogeneous environments.
The fluctuating diffusivity framework allows for a practical analysis of the phenomena of small-molecule diffusion in cementitious materials, where the diffusivity may fluctuate spatiotemporally in response to the heterogeneous nature of the diffusion medium.
Researchers in the field of cement materials should have no difficulties visualizing diffusion phenomena that fall within the scope of such a framework, such as gas diffusion in a depercolated capillary pore network, cases of diffusion coupled with adsorption on the pore wall, or dissolution in the pore solution. Additionally, phenomena such as the consumption of \ce{CO_2} by carbonation and the immobilization of chloride ions through Friedel's salt and calcium oxychloride formation \cite{Suryavanshi1996, Brown2004} may also fall within the scope of this framework if these phenomena are regarded as trapping states for a long time.
When the timescale of observation is comparable to a timescale where the molecules diffuse over the characteristic length of the heterogeneous environment, non-Gaussian behavior of the displacement distribution is exhibited; that is, the tails of the displacement distribution tend to be heavy. In other words, non-Gaussianity can be manifested on the timescale during the diffusion process through the representative elementary volume (REV) in cementitious materials \cite{Ukrainczyk2014, Gao2021}.

Herein, we present several sophisticated approaches for investigating diffusion in cementitious materials.
There are two primary methods for understanding the diffusion phenomena of small molecules in cementitious materials: (i) numerical diffusion simulations of virtual microstructures that replicate the microstructural characteristics of cementitious materials and (ii) empirical or semi-empirical modeling of effective diffusion coefficients through a process of homogenization.
In recent years, the former approach of numerical diffusion simulations on virtual microstructures has made significant progress, effectively simulating the diffusion of various diffusants in cementitious materials of various types and compositions, both with and without transition zones \cite{Zhang2011, Liu2015, Ma2015, Walther2016, Liu2018, Liu2019, Liu2020CCR, Liu2020CCC, Liu2022, Zhang2021}.
A particularly effective recent approach within this model has been the implementation of numerical diffusion models, such as those based on the lattice Boltzmann method \cite{Walther2016, Liu2020CCR, Liu2020CCC, Liu2022}, random walk method \cite{Ma2015, Liu2019, Hailong2021, Zhang2021}, and finite element method \cite{Zhang2011}, which utilize virtual 3D microstructures generated by hydration models.
Several hydration models have been proposed, such as CHEMHYD3D \cite{Bentz1991,Garboczi1992,Bentz2005}, HYMOSTRUC3D \cite{VanBreugel1995}, THAMES \cite{Bullard2011,Feng2014}, DuCOM \cite{Maekawa2003}, IPKM \cite{Pignat2005}, $\mu$ic \cite{Bishnoi2009}, which are widely used in cement and concrete research.
In such microstructure-guided diffusion models, the CHEMHYD3D model (a voxel-based approach) devised by Bentz and Garboczi \cite{Bentz1991,Garboczi1992,Bentz2005} and HYMOSTRUC3D (a vector-based approach) developed by van Breugel  \cite{VanBreugel1995} are commonly utilized \cite{Zhang2011,Zhang2012,Liu2020CCR,Liu2020CCC}.
Both CHEMHYD3D and HYMOSTRUC3D are based on Jennings's colloidal model of Calcium-Silicate-Hydrates (CSH) morphology  \cite{Jennings2000}.
Recently, advancements in the force field of molecular dynamics in cementitious materials have become well developed \cite{Pellenq2009,Mishra2017,Valavi2022}. Zhang et al. modeled diffusion simulations using the random walk method on structures generated by molecular dynamics \cite{Zhang2021}.
The latter approach (ii) describes mass diffusion phenomena through empirical or semi-empirical modeling of the effective diffusion coefficient in heterogeneous media and solving the standard diffusion equations utilizing the effective diffusion coefficient.
The effective diffusion coefficients are modeled under homogenization procedures commonly utilized in composite materials and are inferred to agree with experimental observations and structural insights gained from hydration models \cite{Oh2004, Damrongwiriyanupap2017, Patel2018, Chidiac2019, Achour2020, Yang2020, Shafikhani2020, Gu2022}.
In the realm of finite-element-based analysis utilizing REV meshes (where the discretizing mesh size is generally more significant than the discretization scale in microstructure-guided models), an identical homogenization procedure is applied to assign an effective diffusion coefficient to each REV mesh \cite{Ukrainczyk2014, Gao2021}.
The empirical relationship linking the parameters of the capillary pores and the effective diffusion coefficient is well organized in a critical review article by Patel et al. \cite{Patel2016}.
When the porosity is known, the primary strategy is to attempt to express the effective diffusion coefficient through Archie's law \cite{Archie1942}; when porosity data are unavailable, the effective diffusion coefficient is frequently derived via the Powers model \cite{Brownyard1946}, which can link the hydration degree and water-cement ratio (w/c) to the capillary porosity.
Yamaguchi et al. refined the empirical relationship by assessing accessible capillary pores and demonstrated that the modified model is effective in describing the effective diffusion coefficient of tritiated water \cite{Yamaguchi2009}.
Furthermore, the empirical effective diffusive coefficient was adapted to include semi-empirical parameters that characterize the morphology of the pore network (tortuosity, connectivity, constrictivity, and formation factor) \cite{Promentilla2009, Patel2016, Hatanaka2017, Yio2019, Yang2020, Gao2021}.
Extensive research has been conducted to relate these parameters to the pore topology obtained from imaging techniques rather than simply adjusting bulk diffusion coefficients to effective diffusion coefficients \cite{Lu2006, Wong2006, Promentilla2009, Ukrainczyk2014, Yio2017, Yio2019, Gao2021}.
Recently, an attempt was made to construct a regression model for the diffusion of chloride ions in concrete using machine learning techniques \cite{Liu2021}.
It is important to note that none of the models presented in this paragraph expressing diffusion coefficients can be considered universally applicable.
For instance, the microstructure-based diffusion model in dry cement paste established by Liu et al., despite considering various factors related to multi-scale properties, must explain the diffusion coefficient in low-w/c mixing cement pastes perfectly \cite{Liu2020CCR}.
This discrepancy may be attributed to the structural fluctuations of the generated virtual microstructures, which significantly impact the apparent diffusivity in the low w/c regime.
Additionally, the empirical model appears to exhibit a greater discrepancy between the predicted diffusion coefficients and those observed in the low w/c regime \cite{Patel2016}.
In this study, we introduce an up-to-date theoretical physics concept, ``fluctuating diffusivity," to the cement and concrete fields.
The proposed framework enables incorporating morphological features of heterogeneous media and considering several types of diffusion as stochastic processes without requiring detailed structural information or multiple empirical parameters.

The remainder of this study is structured as follows.
Section 2.1 presents a comprehensive formulation of the fluctuating diffusivity with arbitrary discrete states. In Section 2.2, we delve into a two-state fluctuating diffusivity (2SFD) model following the work of Uneyama et al. \cite{uneyama2019relaxation}, and Miyaguchi et al. \cite{Miyaguchi2019}. We analytically calculated the self-part of the intermediate scattering function and the second and fourth moments of the probability density of the particle displacement.
Section 2.3 describes the application of the 2SFD model for the diffusion of \ce{O_2} in cement pastes under standard temperature and pressure conditions as a preliminary test case.
Section 3 discusses the distinctions of the proposed model in comparison with existing models, its scope of applicability and limitations, its potential for generalization to cementitious systems, and the potential impact of non-Gaussian diffusion on the long-term durability assessment of future structures.
Finally, conclusions are provided in Section 4.

\section{Theory}

\subsection{Fluctuating diffusivity with $n$-states}

The diffusion equation can represent the fluctuating diffusivity \cite{Miyaguchi2019, jain2017diffusing}, which includes a fluctuating diffusivity term $D(t)$, as Eq. \eqref{eq:diffusion_equation_with_fd}.
While this work analyzes the 2SFD model in the following subsections, the calculation method is not restricted to the two states. Thus, we calculate the general $n$-state case as
\begin{equation}
D(t)=\bm{D}^{\top}\bm{\xi}(t)
\label{eq:diffusion_coefficient}
\end{equation}
where $\bm{D}^{\top}=(D_1, D_2,\cdots, D_n)$ is the vector of the diffusion coefficients and its component $D_i$ denotes the diffusion coefficient of the $i$-th state. $\bm{\xi}(t)$ indicates the state of diffusivity at time $t$, and if the diffusivity is in the p-th state at time $t$, each component of $\bm{\xi}(t)$ becomes $\xi_p=1$ and $\xi_{i\neq p}=0$.

Here, we describe the probability density vector, where the particle is in $i$-state at time $t$ as $\bm{P}(t)$, and its stochastic process is described as
\begin{equation}
    \frac{\partial \bm{P}(t)}{\partial t}
    =\bm{R}\bm{P}(t)
    \label{eq:transition_probability}
\end{equation}
where $\bm{R}$ denotes the transition matrix.
From this expression, the probability density of $\bm{P}(t+\Delta)$ with an infinitesimal time step $\Delta$ for a given $\bm{P}(t)$ can be formally expressed as
\begin{equation}
    \bm{P}(t+\Delta)=\exp\left(\Delta \bm{R}\right)\bm{P}(t)
\end{equation}
From this expression, the transition probability, where the state changes from $\bm{\xi}(t)$ to $\bm{\xi}(t+\Delta)$ is
\begin{equation}
    P(\bm{\xi}(t+\Delta); \bm{\xi}(t))=
    \bm{\xi}^\top(t+\Delta)
    \exp\left(\Delta\bm{R}\right)
    \bm{\xi}(t)
\end{equation}

To proceed with the calculation of Eq. ~\eqref{eq:diffusion_equation_with_fd}, intermediate scattering function $F(\bm{k}, t)=\int e^{-i\bm{k}\cdot \bm{r}} G(\bm{r}; t)$ is useful.
By taking the Fourier transform of Eq. ~\ref{eq:diffusion_equation_with_fd}, we obtain the differential equation for $F(\bm{k}, t)$ as follows:
\begin{equation}
    \frac{\partial F(\bm{k}; t)}{\partial t}
    =-D(t) k^2 F(\bm{k}; t).
\end{equation}
This differential equation is formally solved as  \cite{uneyama2015fluctuation, miyaguchi2017elucidating}
\begin{equation}
    F(\bm{k}; t)=\left\langle
    \exp\left(-k^2\int_0^t D(t') dt'\right)\right\rangle_D
    \label{eq:isf_formal}
\end{equation}
where $\langle \cdots \rangle_D$ denotes the ensemble average $D(t)$.
Formally, Eq.~\eqref{eq:isf_formal} can be described in discretized form as follows:
\begin{equation}
\begin{split}
    F(\bm{k}; t)
    =&\sum_{\bm{\xi}(j\Delta t)}\exp\left[-\sum_{j=0}^{t/\Delta -1}\Delta k^2 \bm{D}^\top\bm{\xi}(j\Delta)\right]\times\\
    &\prod_{j=0}^{t/\Delta -1}
    \left[
    P(\bm{\xi}((j+1)\Delta); \bm{\xi}(j\Delta))
    \right]
    \bm{\xi}^\top(0)\bm{P}(0)\\
    =&\sum_{\bm{\xi}(j\Delta)}
    \prod_{j=0}^{t/\Delta -1}
    \left\{
    \exp\left[
    -\Delta k^2 \bm{D}^\top\bm{\xi}(j\Delta)
    \right]\times\right.\\
    &\left.\bm{\xi}^\top((j+1)\Delta)
    \exp\left(\Delta\bm{R}\right)
    \bm{\xi}(j\Delta)
    \right\}
    \bm{\xi}^\top(0)\bm{P}(0)
\end{split}
\label{eq:isc_explicit}
\end{equation}
This equation is similar to the partition function of the Ising model under an external field.
Then, we define the transfer matrix $\bm{T}$ as
\begin{equation}
\begin{split}
    \bm{\xi}^\top \bm{T} \bm{\xi}(t)
    =&
    \bm{\xi}^\top(t+\Delta)
    \exp\left[\Delta\bm{R}
    -\frac{\Delta k^2 \bm{D}^\top\
    [\bm{\xi}(t+\Delta)+\bm{\xi}(t)]}{2}
    \right]
    \bm{\xi}(t)
\end{split}
    \label{eq:transfer_matrix}
\end{equation}
Because $\Delta$ is an infinitesimal quantity, the elements of the transfer matrix can be expressed as
\begin{equation}
T_{ij}=\exp(\Delta \bm{R})_{ij}\exp(-\Delta k^2 D_{j} \delta_{ij})=\delta_{ij}+\Delta(R_{ij}-k^2D_{j}\delta_{ij})
\end{equation}
For brevity, we define matrix $Q_{ij}$ as
\begin{equation}
T_{ij}=\delta_{ij}+\Delta Q_{ij}
\label{eq:matrix_exponent}
\end{equation}

Utilizing the transfer matrix, Eq. ~\eqref{eq:isc_explicit} can be reduced to:
\begin{equation}
\begin{split}
    F(\bm{k}; t)
    =&\sum_{\bm{\xi}(j\Delta)}
    e^{\Delta k^2\bm{D}^\top\bm{\xi}(t)/2}\times\\
    &\prod_{j=0}^{t/\Delta -1}
    \bm{\xi}^\top((j+1)\Delta)\bm{T}
    \bm{\xi}(j\Delta)
    e^{-\Delta k^2\bm{D}^\top\bm{\xi}(0)/2}
    \bm{\xi}^\top(0)\bm{P}(0)\\
    =&\sum_{\bm{\xi}(j\Delta)}
    \prod_{j=0}^{t/\Delta -1}
    \bm{\xi}^\top((j+1)\Delta)\bm{T}
    \bm{\xi}(j\Delta)
    \bm{\xi}^\top(0)\bm{P}(0)\\
    =&\sum_{\bm{\xi}(t)}
    \bm{\xi}^\top(t)
    \bm{T}^{t/\Delta}\bm{P}(0)
    =\sum_{\bm{\xi}(t)}
    \bm{\xi}^\top(t)
    e^{t\bm{Q}}\bm{P}(0)
\end{split}
\label{eq:isf_general}
\end{equation}
This equation can be calculated when the initial probability density $\bm{P}(0)$, $i$-th state diffusivity coefficient from Eq. \eqref{eq:diffusion_coefficient}, and the transition probability $\bm{R}$ from Eq.\eqref{eq:transition_probability} are provided.

\subsection{2SFD model}

Here, we consider the two-state fluctuating diffusivity (2SFD) model following the literature by Uneyama et al. \cite{uneyama2019relaxation}, and Miyaguchi et al. \cite{Miyaguchi2019}, which serves as a mathematically tractable model.
The diffusivity of the particle in the 2SFD model is characterized by distinct variables $\bm{D}^{\top}=(D_f, D_s)$ and the transition probability matrix $\bm{R}$, which is represented as
\begin{equation}
    \bm{R}=\begin{pmatrix}
    -\omega_f && \omega_s \\
    \omega_f && -\omega_s
    \end{pmatrix}
\end{equation}
In the equilibrium state, the initial probability density is given by:
\begin{equation}
    \bm{P}(0)=
    \frac{1}{\omega_f+\omega_s}
    \begin{pmatrix}
    \omega_s\\ \omega_f
    \end{pmatrix}
\end{equation}
Then, matrix $\bm{Q}$ in Eq. ~\eqref{eq:isf_general} is expressed as
\begin{equation}
    \bm{Q}=\begin{pmatrix}
        -\omega_f-k^2 D_f && \omega_s\\
        \omega_f && -\omega_s-k^2 D_s
    \end{pmatrix}
    \label{matrixq}
\end{equation}
For this $\bm{Q}$, the eigenvalues and corresponding eigenvectors are given by
\begin{align}
    \lambda_{\pm}&=-\frac{\omega_f+k^2D_f+\omega_s+k^2D_s\pm \sqrt{(\omega_f+k^2D_f-\omega_s-k^2D_s)^2+4\omega_f\omega_s}}{2}
    \label{eq:eigenvalue}\\
    \bm{v}_{\pm}&=\begin{pmatrix}
        -\frac{\omega_f+k^2D_f-\omega_s-k^2D_s\pm \sqrt{(\omega_f+k^2D_f-\omega_s-k^2D_s)^2+4\omega_f\omega_s}}{2\omega_f}\\
        1
    \end{pmatrix}
    \label{eq:eigenvector}
\end{align}
Using $\lambda_{\pm}$ and $\bm{v}_{\pm}$, matrix $\bm{Q}$ can be described as
\begin{equation}
    \bm{Q}=(\bm{v}_+, \bm{v}_-)
    \begin{pmatrix}
        \lambda_+ && 0\\
        0 && \lambda_-
    \end{pmatrix}
    (\bm{v}_+, \bm{v}_-)^{-1}
    \label{eq:diagonal}
\end{equation}
Combining Eq.~\eqref{eq:isf_general} and \eqref{eq:diagonal}, we obtain:
\begin{equation}
\begin{split}
    F(\bm{k}; t)=&(1,1)
    (\bm{v}_+, \bm{v}_-)
    \begin{pmatrix}
        e^{\lambda_+ t} && 0\\
        0 && e^{\lambda_- t}
    \end{pmatrix}
    (\bm{v}_+, \bm{v}_-)^{-1}
    \frac{1}{\omega_f+\omega_s}\begin{pmatrix}
        \omega_s\\\omega_f
    \end{pmatrix}\\
    =&\chi_+ e^{\lambda_+ t} + \chi_- e^{\lambda_- t}
\end{split}
\label{eq:isf_result}
\end{equation}
where we defined $\chi_{\pm}$ as
\begin{equation}
    \chi_{\pm} =
    \frac{1}{2}\left[
    1\pm \frac{(k^2D_f-k^2D_s)(\omega_f-\omega_s)+(\omega_s+\omega_f)^2}{(\lambda_+ - \lambda_-)(\omega_f+\omega_s)}\right]
\end{equation}
Eq.~\eqref{eq:isf_result} includes all information for the probability density $G(\bm{r}, t)$.
From Eq.~\eqref{eq:isf_result}, we calculate all moments of the probability density, such as the second and fourth moments ($\langle \bm{r}^2(t)\rangle$ and $\langle \bm{r}^4(t)\rangle$), where the bracket $\langle \cdots\rangle$ denotes the statistical average.
The utilization of higher moments quantifies the deviation of $G(\bm{r}; t)$ from the Gaussian distribution, as discussed subsequently.
According to the definition of the self-part of the intermediate scattering function, these moments are formally obtained in an isotropic system as follows:
\begin{align}
    &\langle \bm{r}^2(t)\rangle
    =-\frac{\partial^2}{\partial \bm{k}^2}F(\bm{k}, t)|_{k=0}
    \label{eq:moment2_formal}\\
    &\langle \bm{r}^4(t)\rangle
    =\frac{\partial^2}{\partial \bm{k}^2}
    \frac{\partial^2 F(\bm{k}, t)}{\partial \bm{k}^2}|_{k=0}
    \label{eq:moment4_formal}
\end{align}
To assign Eq.~\eqref{eq:isf_result} to Eq.~\eqref{eq:moment2_formal}, we obtain:
\begin{align}
    \langle \bm{r}^2(t)\rangle
    =6\frac{D_f\omega_s+D_s\omega_f}{\omega_f+\omega_s}t ,
    \label{eq:moment2_result}
\end{align}
Using this relation, the average diffusion coefficient $D$ can be determined through the relation $\langle \bm{r}^2(t)\rangle=6Dt$ in a three-dimensional system, asfollows:
\begin{equation}
    D=\frac{D_f\omega_s+D_s\omega_f}{\omega_f+\omega_s}
    \label{eq:diffusion_coefficient_result}
\end{equation}
This outcome indicates that the average diffusion coefficient in the present 2SFD model is the weighted average of $D_f$ and $D_s$ with transition rates $\omega_f$ and $\omega_s$.
Furthermore, by utilizing Eq. \eqref{eq:moment4_formal}, we obtain an analytical expression for the fourth moment of $G(\bm{r}; t)$ as
\begin{equation}
\begin{split}
    \langle \bm{r}^4(t)\rangle
    =&120\left\{\frac{(D_f\omega_s+D_s\omega_f)^2}{2(\omega_f+\omega_s)^2}t^2-\right.\\
    &\left.\frac{(D_f-D_s)^2\omega_f\omega_s}{(\omega_f+\omega_s)^4}
    \left[1-(\omega_f+\omega_s)t-e^{-(\omega_f+\omega_s)t}\right]\right\},
\end{split}
    \label{eq:moment4_result}
\end{equation}
which is used later.

\subsection{Application of 2SFD model to gas \ce{O_2} in cement pastes}
\begin{figure}
    \centering
    \includegraphics[width=0.8\textwidth]{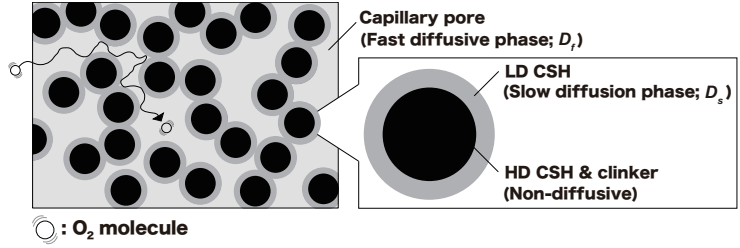}
    \caption{Schematic figure of \ce{O_2} diffusion in a cement paste, consisting of three phases: capillary pore, low-density CSH (diffusive) phase, and non-diffusive phase (high-density CSH and unhydrated cement clinker)}
    \label{fig:system schematic diagram}
\end{figure}

\begin{figure}
    \centering
    \includegraphics[width=0.8\textwidth]{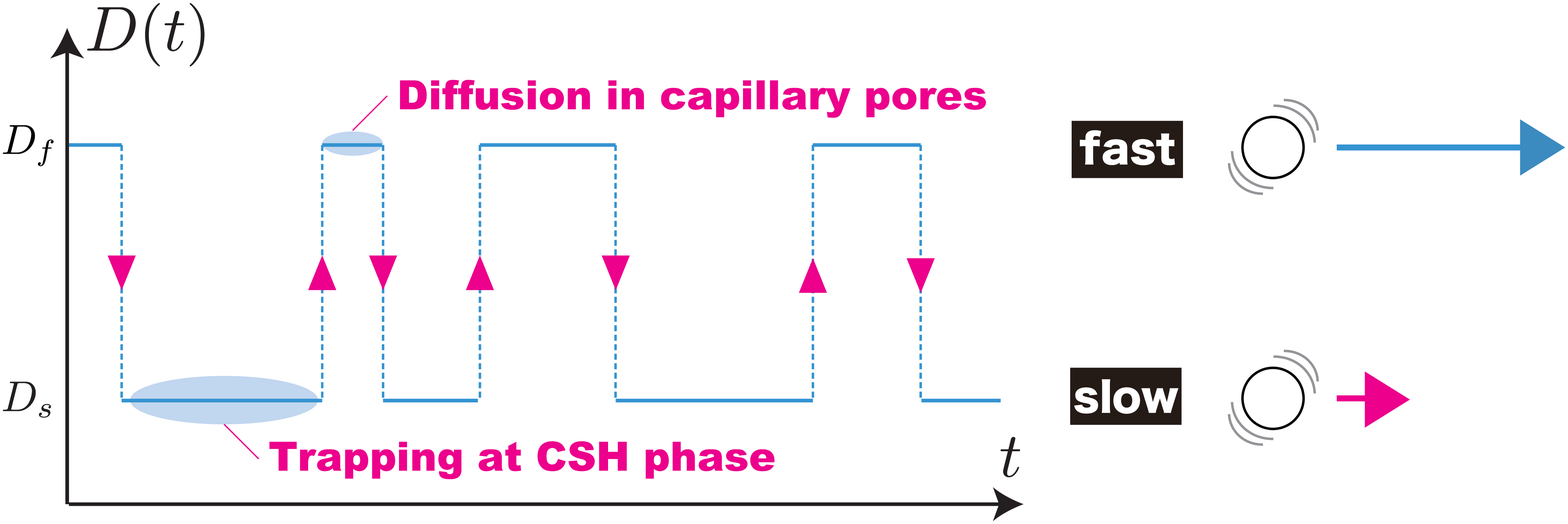}
    \caption{Schematic illustration of
    transitional process of diffusivity $D(t)$. $D_f$ corresponds diffusion coefficient of the fast state in capillary pores, and $D_s$ corresponds diffusion coefficient of the slow state at CSH phase.}
    \label{fig:schematic illustration of 2 state diffusion model}
\end{figure}

In this study, we addressed the fundamental problem of \ce{O_2} diffusion, which is known to be one of the basic aggressive gases that can affect the long-term performance of reinforced concrete structures \cite{Lawrence1984}.
The diffusion of oxygen in the dry cement paste (i.e., the absence of free water in capillary pores) was chosen as the primary case study.
This system was selected because it presents a relatively simple diffusion medium for cementitious materials but offers some heterogeneity.
Here, we focus on the \ce{O_2} diffusion in dry cement paste consisting of the capillary pore phase and colloidal CSH \cite{Jennings2000} phase under ambient temperature and pressure conditions $T=298$ K and $P=1$ atm.
As depicted in Figure \ref{fig:system schematic diagram}, colloidal CSH consists of two different density phases in proximity to the surface and hydration front, which are classified as LD-CSH (low-density CSH) and HD-CSH (high-density CSH), respectively, \cite{Tennis2000,Jennings2007}.
For simplicity, this study treats the capillary pore and LD-CSH phases as diffusive, while the HD-CSH and unhydrated clinker regions as non-diffusive phases.
Given the non-negligible difference in density between LD-CSH and HD-CSH, we tentatively assumed that \ce{O_2} molecules could not penetrate the HD-CSH phase through the LD-CSH phase.
This study regards the diffusion in the capillary void as a ``rapid'' diffusion process (diffusion coefficient $D_f$), comprising both molecular and Knudsen diffusion. In contrast, diffusion in the LD-CSH is considered a ``slow'' process (diffusion coefficient $D_s$).
They are used as inputs for the 2SFD model, as illustrated in Figure \ref{fig:schematic illustration of 2 state diffusion model}.
The following analyses derive all heterogeneous diffusion media characteristic values through physically reasonable estimations.
In our system, at ambient temperature and pressure, the impact of surface diffusion on the overall diffusion characteristics is possibly negligible (the coverage of the \ce{O_2} molecule is approximately $0.01$ or less, which could be estimated similarly as in Ref. \cite{Liu2020CCR}).

From this point onward, the system setup is described in detail.
The size of the colloidal CSH was assumed to be $l=50$ nm, which was determined based on the size of the globule flocs in the CM-II model proposed by Jennings \cite{Jennings2008}.
In this study, the thickness of the LD-CSH on colloidal CSH, which is treated as the diffusive phase, is assumed to be $10$ nm from the surface following the value utilized in the previous microstructure-guided model \cite{Liu2020CCR}.
For simplicity, we assume that the colloidal CSH is spherical and the number density of the colloidal CSH is denoted by $\rho$.
Porosity is represented by $\phi$, which can also be represented by $\rho$ and $l$ as
\begin{equation}
    1-\phi=\frac{\rho \pi l^3}{6}
    \label{eq:relation_porosity_density}
\end{equation}

Using the parameters specified above, we describe the four input parameters, namely $D_f$, $D_s$, $\omega_f$, and $\omega_s$, in the 2SFD model.
In the present model, the diffusion coefficient of the fast state $D_f$ can be considered as the harmonic average of the molecular and Knudsen diffusion coefficients $D_M$ and $D_K$, as follows:
\begin{equation}
    D_f=\frac{D_MD_K}{D_M+D_K}
    \label{eq:df_harmonic_average}
\end{equation}
Under ordinary pressure and temperature conditions, $D_M$ is estimated as \cite{chapman1990}
\begin{equation}
    D_M=\frac{3k_BT}{8 P \sigma^{2}}\sqrt{\frac{k_BT}{\pi m}}
    \label{eq:moleculr_diffusion}
\end{equation}
where $k_B$ denotes the Boltzmann constant.
$\sigma$ and $m$ represent the diameter and mass of the oxygen, respectively, and
they are given by $\sigma=3.46\times 10^{-10}m$ \cite{koo2017accelerating, lv2019h2} and $m=5.31\times 10^{-26}kg$, respectively.
From these variables, $D_M$ is estimated to be $D_M=1.99\times 10^{-5}m^2s^{-1}$.
In complex systems such as cement materials, estimating $D_K$ is difficult.
We roughly estimated $D_K$ by approximating the structure of the target cement system as a Lorentz gas, that is, a single mobile particle in fixed spherical obstacles.
An analogous postulation was utilized in the research examining gas diffusion in cement paste by Liu et al. \cite{Liu2020CCR}.
Under this assumption, the diffusion coefficient was determined as \cite{dorfman2021}
\begin{equation}
    D_K=\frac{\bar{v}^2\tau}{3}
    \label{eq:dk}
\end{equation}
where $\bar{v}$ denotes the mean speed of oxygen, given by $\bar{v}=\sqrt{8k_BT/\pi m}$, and $\tau$ represents the mean free time.
Estimating $\tau$ is challenging; however, it has been roughly estimated from the mean pore size \cite{Liu2020CCR}.
This study obtained a rough approximation of $\tau$ by considering gas kinetics.
When the colloidal CSH is diluted, the mean free time can be expressed as $\tau=4/\rho\pi l^2 \bar{v}$, where it is assumed that the interaction distance between \ce{O_2} and colloidal CSH is approximated as $(l+\sigma)/2 \simeq l/2$.
The estimated $\tau$ is inadequate for the low-porosity regime; for instance, $\tau$ should be $0$ for $\phi=0$.
To account for the case of a small $\phi$, a phenomenological description of $\tau$ as depicted in the previous literature \cite{dellago2001field} is employed:
\begin{equation}
    \tau=\left(1-\frac{\rho\pi l^3}{6}\right)\frac{4}{\rho\pi l^2 \bar{v}}
    \label{eq:freetime_knudsen}
\end{equation}
Combining Eqs.~\eqref{eq:relation_porosity_density}, \eqref{eq:dk}, and \eqref{eq:freetime_knudsen}, we obtain
\begin{equation}
    D_K=\frac{4l\phi}{9(1-\phi)}\sqrt{\frac{2k_BT}{\pi m}}
    \label{eq:knudsen_diffusion}
\end{equation}
In this expression, $D_K$ becomes $0$ for $\phi=0$ and diverges for $\phi=1$, which agrees with the intuitive representation of Knudsen diffusion.

The slow diffusion state pertains to the diffusion within the LD-CSH phase.
The diffusivity determination is not straightforward because handling diffusion within the LD-CSH phase is complex.
Although this estimation remains an open problem, prior investigations suggest that there may exist two possible approaches: (i) consider it as surface diffusion and determine the diffusion coefficient through Wu's empirical equation \cite{Wu2015} and the model of Chen and Yang \cite{Chen1991}, which are commonly employed in the context of shale gas, or (ii) utilize effective medium theory, as demonstrated by Patel et al. \cite{Patel2018}.
Here, we assumed a slow diffusion coefficient of $D_s=10^{-8}m^2s^{-1}$.
This value was consistent with the estimations described above.
The first approach necessitates isosteric adsorption heat ($\Delta H$) as input for Wu's empirical equation \cite{Wu2015}. If we adopt the isosteric adsorption heat of \ce{CO_2} on the CSH surface, $\Delta H \sim 10$ kJ/mol is tentatively applied to \ce{O_2} using the same procedure conducted by Liu et al. \cite{Liu2020CCR}, and $D_s$ would be on the order of $10^{-8} m^{2}s^{-1}$.
Furthermore, Patel et al. reported that the C-S-H diffusivity was three orders of magnitude lower than the bulk diffusivity for various diffusants \cite{Patel2018}.
Subsequently, the transition rates $\omega_f$ and $\omega_s$ were determined consistently with the information on the pore structure.
$\omega_f$ corresponds to the transition rate from the fast diffusion state in the capillary pores to the slow diffusion state in the LD-CSH phase.
In the dilute limit of the volume fraction of the CSH phase, the average capillary pore size $L$ can be approximated as $\rho^{-1/3}$.
When the volume fraction of the CSH phase is not diluted, the effect of the excluded volume must be considered, which can be phenomenologically estimated.
As $L$ approaches $0$ when the space is entirely occupied by the CSH phase and diverges when CSH is absent, a possible relation between $L$ and the CSH number density is
\begin{equation}
 L\simeq\left[\frac{\rho}{(1-\rho\pi l^3/6)}\right]^{-1/3}
 =l\left[\frac{\pi\phi}{6(1-\phi)}\right]^{1/3}
\end{equation}
If it is assumed that $\omega_f$ represents the rate of contact with colloidal CSH from the capillary pore phase, $\omega_f$ can be estimated as
\begin{equation}
    D_f\simeq L^2\omega_f
    \label{eq:relation_df_rf}
\end{equation}
$\omega_s$ represents the transition rate from the slow diffusion state in the LD-CSH phase to the fast diffusion state in the capillary pores.
Considering that the thickness of the LD-CSH phase is approximately $l_{LD}=10$ nm and that it can escape from the surface by approximately $10$ nm motion, the following estimation can be obtained:
\begin{equation}
    D_s\simeq l_{LD}^2 \omega_s
    \label{eq:relation_ds_rs}
\end{equation}
By utilizing the relations $D_f$, $D_s$, $\omega_f$, and $\omega_s$ as stated above, we can calculate the dynamics of \ce{O_2} in the 2SFD model.

\begin{figure}
    \centering
    \includegraphics[width=0.6\textwidth]{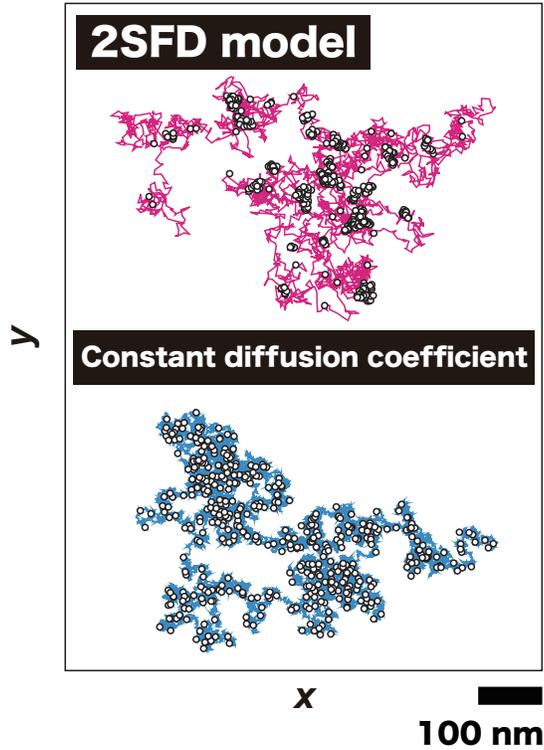}
    \caption{The trajectory of $O_2$ over the observed time duration $1000/(\omega_f+\omega_s)$ at a porosity $\phi=0.5$ is represented by the pink curve. For comparison, the blue curve depicts the trajectory of particle diffusion with a constant diffusion coefficient, as represented by equation (\ref{eq:diffusion_coefficient_result}). Closed circle symbols are also displayed at the same time intervals of $(\omega_f+\omega_s)^{-1}$.}
    \label{fig:trajectory}
\end{figure}

The representation of the trajectory may provide an intuitive understanding of the 2SFD model. Subsequently, using a kinetic Monte Carlo scheme  \cite{gillespie1976general, bortz1975new} based on Equations~\eqref{eq:diffusion_equation_with_fd} and \eqref{eq:transition_probability}, we numerically calculated the trajectory. Figure~\ref{fig:trajectory} illustrates a representative trajectory of an \ce{O_2} molecule in the 2SFD model, depicted by a pink curve with closed circles, indicating a time interval of $(\omega_f+\omega_s)^{-1}$. For comparative purposes, the trajectory of an \ce{O_2} molecule moving without fluctuating diffusivity (the diffusion coefficient remains constant as per Equation~\eqref{eq:diffusion_coefficient_result}) is represented by a blue curve with closed circles plotted at every time interval $(\omega_f+\omega_s)^{-1}$. The pink curve effectively captures the heterogeneous diffusivity, which can be interpreted as a reflection of the heterogeneous nature of the cement paste. In contrast, the blue curve does not exhibit heterogeneity.

\begin{figure}[htb]
    \centering
    \includegraphics[width=0.8\textwidth]{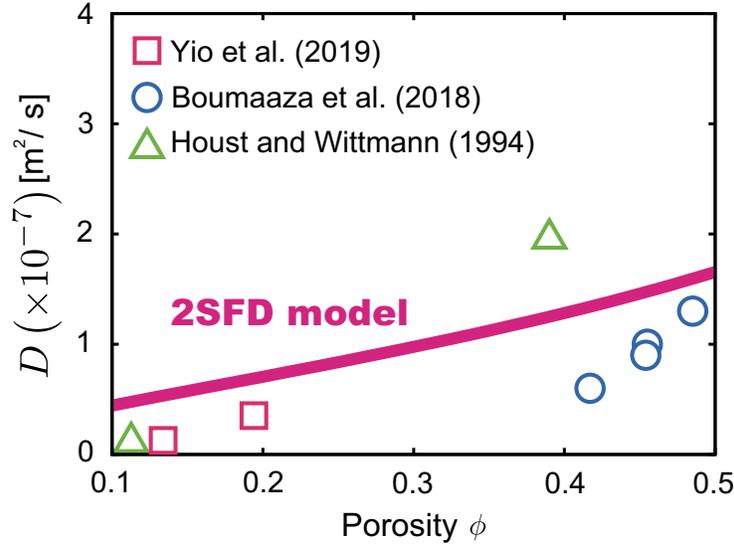}
    \caption{Diffusion coefficient of the molecule \ce{O_2} against the porosity $\phi$. For comparison, the data reported by Yio et al. \cite{Yio2019}, Boumaaza et al. \cite{Boumaaza2018}, and Houst and Wittmann \cite{Houst1994} are also displayed.}
    \label{fig:diffusion_coefficiet_porossity}
\end{figure}

From Equation (\ref{eq:diffusion_coefficient_result}), we depict the diffusion coefficient $D$ as a function of the porosity $\phi$ in Figure (\ref{fig:diffusion_coefficiet_porossity}). For comparative purposes, data obtained in previous studies by Yio et al. \cite{Yio2019} and Boumaaza et al.  \cite{Boumaaza2018}, and Houst and Wittmann \cite{Houst1994} are represented by blue, purple, and green symbols, respectively.
Table \ref{datatable} summarizes the detailed conditions of previous studies that measured the oxygen diffusion coefficients in cement paste.
In this study, the ideal comparison for the measured \ce{O_2} diffusivity in cement pastes was based on data obtained from completely dry cement pastes, as reported by Boumaaza et al. \cite{Boumaaza2018}. However, to the best of our knowledge, such data are limited. Therefore, to provide a reasonable comparison, we elected to include the results of previous studies that have measured \ce{O_2} diffusivity in cement pastes under conditions of relatively low humidity, as suggested by the findings of Houst and Wittmann \cite{Houst1994} that the effect of relative humidity on diffusivity is minimal below 55\%. Specifically, we included the results of Yio et al. \cite{Yio2019}, and Houst and Wittmann \cite{Houst1994} as comparable data for the \ce{O_2} diffusivity in cement pastes.
However, we did not include the results of Yio et al. \cite {Yio2019}, in which the hydration reaction still needed to be completed in the comparison data. The size of the colloidal CSH changes as the hydration reaction progresses, which affects the estimated diffusion coefficient in this 2SFD model, as discussed later in the study (see Discussion section).
Our theoretical results are in qualitative agreement with the data presented in prior works. It is important to note that the four inputs $D_f$, $D_s$, $\omega_f$, and $\omega_s$ are derived from the system parameters and can be determined through physical considerations.

\begin{table}
    \centering
    \caption{Detail information of previous gas diffusion datasets.}
    \label{datatable}
    \begin{center}
        \scalebox{0.6}{
        \begin{tabular}{c c c c c c} \hline
            Ref. & w/c ratio & Curing & Drying method & Porosity & Conditions \\
        \hline
            Yio et al. \cite{Yio2019} &0.30 & Cured at 100\% RH,& Kept in 55\% RH, 293 K & 0.133 & $0.5 \sim 2.5$ atm\\
            &0.45&293 K for 90 days& & 0.194 &and room temperature\\
            Boumaaza & 0.50 & Cured at 100\% RH & Oven-dried & 0.492 & 1 atm and 293 K \\
            et al. \cite{Boumaaza2018} & & for 1 day, 2 months and 8 month & & 0.455 & \\
             & & & & 0.417 & \\
             & 0.60 & & & 0.483 & \\
             & & & & 0.454 & \\
             Houst and & 0.40 & Immersed in lime water & Oven-dried & 0.110 & 1 atm, room temperature\\
             Wittmann \cite{Houst1994} & 0.80 & for 6 months or more &  & 0.390 & and 47 \% RH \\ \hline
        \end{tabular}
        }
    \end{center}
\end{table}

\begin{figure}
    \centering
    \includegraphics[width=0.8\textwidth]{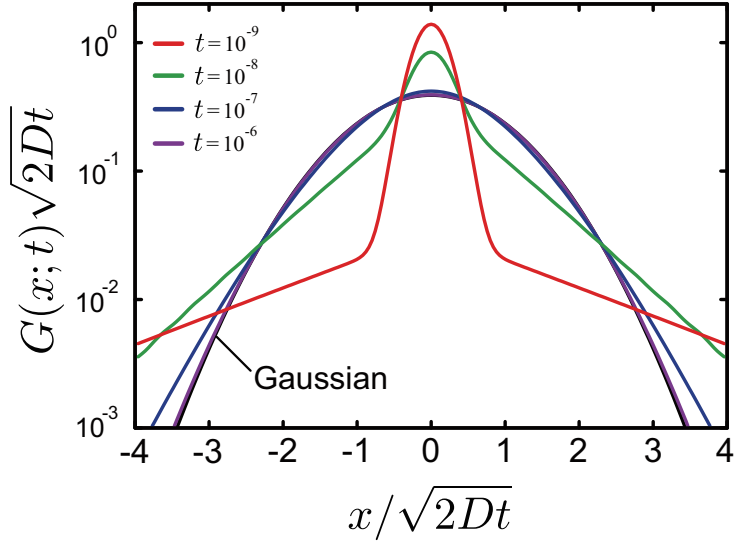}
    \caption{Probability density of $O_2$ displacement for various time duration $t$ at the porosity $\phi=0.5$. The horizontal and vertical axes are normalized by the standard deviation $\sqrt{2Dt}$. For comparison, the Gaussian distribution is also presented with the black curve.}
    \label{fig:van-hove}
\end{figure}

The diffusion coefficient often characterizes gas diffusion in cementitious materials; however, more is needed to fully explore concrete structures' reliability. The tail of the probability density of displacement $G(\bm{r};t)$ should also be considered.
Herein, we analyze the probability density of the displacement for a single degree of freedom $x$, $G(x;t)$, which is derived from the inverse Fourier transform of the self-part of the intermediate scattering function as $G(x; t)=(2\pi)^{-1}\int e^{ik_x x}F(k_x; t)$. $F(k_x; t)$ can be computed using Eq. \eqref{eq:diffusion_equation_with_fd} by substituting $\bm{r}$ with $x$; as a result, Eq.\eqref{eq:isf_result} where $\bm{k}$ is substituted with $k_x$ is obtained.
Because the analytical calculation of the inverse transformation of $F(k_x; t)$ is difficult, we perform numerical integration. Fig.~\ref{fig:van-hove} illustrates the probability density of \ce{O_2} displacement for various time durations $t$ at a typical porosity $\phi=0.5$. We scaled the horizontal and vertical axes using the standard deviation of the displacement $\sqrt{2Dt}$ and displayed the Gaussian distribution function as a reference. In the short timescale $t\le 10^{-8}s$, clear deviations of $G(x; t)$ from the Gaussian distribution are observed. These deviations gradually diminish with increasing observation time $t$. $t\sim 10^{-8}$ s is comparable to the timescale of the inverse of $\omega_f$ or $\omega_s$. This result suggests that the Gaussian approximation for $G(x; t)$ may not be appropriate for the timescale over which \ce{O_2} diffuses the lengths of the colloidal CSH or capillary pore.
Our result is reasonable because non-Gaussian distributions have been frequently observed at the microscopic scale in various heterogeneous systems, such as confined water in CSH  \cite{Qomi2014}, glass-forming liquids \cite{rusciano2022fickian}, and colloidal suspensions \cite{pastore2021rapid, kim2013simulation}.

\begin{figure}
    \centering
    \includegraphics[width=0.8\textwidth]{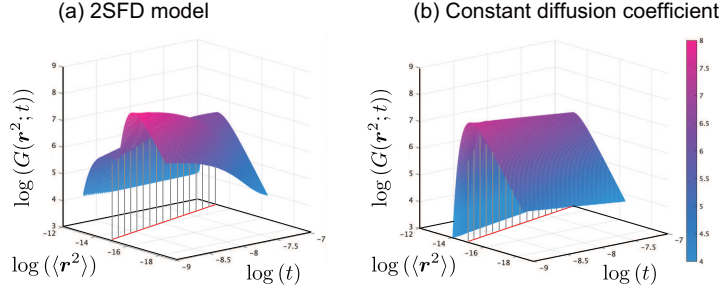}
    \caption{Theoretical result of the probability density and mean square displacement of \ce{O_2} in cement paste obtained from the 2SFD model are shown in Fig.\ref{fig:logloglogplot3d}(a). Fig.\ref{fig:logloglogplot3d}(b) presents the Gaussian distribution and mean square displacement, along with the diffusion coefficient from equation \eqref{eq:diffusion_coefficient_result}, for comparison.}
    \label{fig:logloglogplot3d}
\end{figure}

Representing the mean square displacement $\langle \bm{r}^2(t)\rangle$ with the probability density $G(r^2;t)$ would also be helpful for an intuitive understanding.
Taking the coordinates where $\bm{r}=(0,0,r)$ defines the $z$-axis, we can calculate $G(r^2;t)$ using Eq. ~\eqref{eq:isf_result} as follows:
\begin{equation}
    G(r^2;t)=4\pi r^2 (2\pi)^{-3}\int e^{i\bm{k}\cdot\bm{r}} F(\bm{k}; t) d\bm{k}
\end{equation}
The integration was performed numerically.
Fig.\ref{fig:logloglogplot3d}(a) displays $\langle \bm{r}^2(t)\rangle$ and $G(r^2;t)$ at the porosity $\phi=0.5$.
For comparison, the Gaussian distributions are shown in Fig.\ref{fig:logloglogplot3d}(b).
$\langle \bm{r}^2(t)\rangle$ is a simple linear function of time. However, we observe that $G(r^2;t)$ has a heavy tail compared to the Gaussian distribution over short timescales. This non-Gaussian distribution disappears over a long timescale.
This result visually indicates that the dynamics on a short timescale cannot be described using a Gaussian distribution.

\begin{figure}
    \centering
    \includegraphics[width=0.8\textwidth]{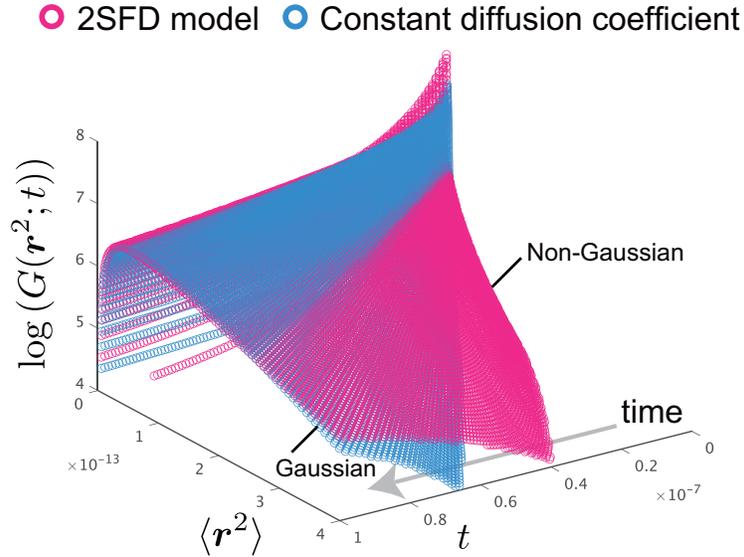}
    \caption{Comparison of probability density with time and displacement, expressed on linear axes, between the 2SFD model (represented by blue open symbols) and a constant diffusion coefficient model (represented by red open symbols. The results in this figure are the same as in Figure \ref{fig:logloglogplot3d}.}
    \label{fig:linearplot3d}
\end{figure}

As a supplement, Figure \ref{fig:linearplot3d} shows the comparison of the probability density with time and displacement, expressed on linear axes, between the 2SFD model and a constant diffusion coefficient model. It presents that the heavy tail observed in the short-time region of the 2SFD model disappears, and the diffusion coefficient becomes asymptotically to the Gaussian distribution over a long timescale. This suggests that the conventional diffusion equation, which attributes an effective diffusion coefficient, may underestimate the probability of long-distance diffusion for a given time scale as long as a heavy tail still exists. This finding highlights the significance of considering non-Gaussian behavior in reliability analysis.

\begin{figure}
\centering
\includegraphics[width=0.8\textwidth]{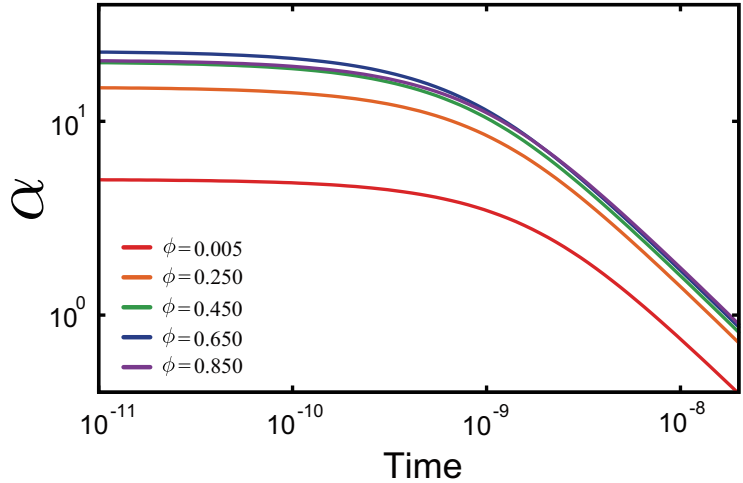}
\caption{Non-Gaussian parameter $\alpha$ against time $t$ with various porosity $\phi$.}
\label{fig:ngp_o2_cs}
\end{figure}

To characterize non-Gaussian diffusion, the non-Gaussian parameter $\alpha$ is often employed \cite{kob1997dynamical,vorselaars2007non,nakai2023fluctuating} and is defined in three-dimensional systems as \cite{rahman1964correlations}
\begin{equation}
\alpha(t)=\frac{3\langle \bm{r}^4(t)\rangle}{5\langle \bm{r}^2(t)\rangle^2}-1
\end{equation}
where brackets denote the statistical averages. $\alpha$ is equal to zero when the stochastic process of displacement conforms to a Gaussian distribution. If the conventional diffusion equation can describe the dynamics of the particle with constant diffusivity, $\alpha$ is equal to zero. Empirically, non-Gaussianity cannot be neglected when $\alpha>0.1$.
We obtain the second moment $\langle \bm{r}^2(t)\rangle$ and fourth moment $\langle \bm{r}^4(t)\rangle$ as represented by Equations~\eqref{eq:moment2_result} and \eqref{eq:moment4_result}, respectively. Consequently, we can determine $\alpha$ as the following expression:
\begin{equation}
    \alpha(t)
    =\frac{2(D_f-D_s)^2\omega_f\omega_s}{(D_f\omega_s+D_s\omega_f)^2(\omega_f+\omega_s)^2}
    \frac{e^{-(\omega_f+\omega_s)t}+(\omega_f+\omega_s)t-1}{t^2}
\end{equation}
Fig.~\ref{fig:ngp_o2_cs} displays $\alpha$ against time $t$ for various porosities $\phi$.
$\alpha$ exhibits strong non-Gaussianity for the small-time regime $t\ll 10^{-9}$, and does not monotonically change with increasing porosity $\phi$.
This result may be reasonable because the heterogeneity of diffusivity disappears for $\phi=0$ and $\phi=1$.
Additionally, the non-Gaussian parameter $\alpha$ decreases from $10^{-9}s<t<10^{-8}$, indicating that diffusion on the timescale of $t<10^{-8}$ cannot be described by a Gaussian process or a conventional diffusion equation with constant diffusivity.

\section{Discussion}
In this study, we employed the 2SFD model for \ce{O_2} diffusion in cement pastes, which constitutes a stochastic diffusion model comprising parameters that can be physically inferred from an abundance of experimental studies on gas diffusivity in cementitious materials while incorporating some crucial aspects of microstructures.
This model effectively addresses stochastic processes involving transitions between multiple diffuse states and can analytically determine the probabilistic displacement distribution, including the non-Gaussian parameter.
Therefore, it constitutes a highly flexible framework that can be easily modified if the transition rates between multiple diffuse states, including additional states, can be effectively assessed.

\begin{figure}
    \centering
    \includegraphics[width=0.8\textwidth]{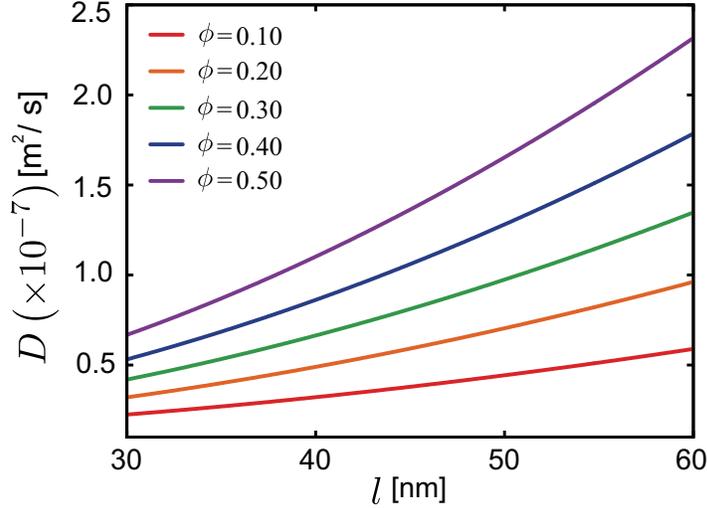}
    \caption{Diffusion coefficient $D$ against the colloidal CSH size $l$ with various porosity $\phi$.}
    \label{fig:d_csh-length}
\end{figure}

We tentatively assumed colloidal CSH dimensions of 50 nm in the above analysis. This is possibly acceptable because Jennings's CSH morphological model of CM-II \cite{Jennings2008} suggests that the size of globule flocs is within the range of 30–60 nm.
As one of the outputs of the 2SFD model, the estimated net diffusion coefficient increased as the assumed colloidal CSH size increased, as shown in Figure \ref{fig:d_csh-length}.
This behavior is consistent with the experimental results reported by Bentz et al. \cite{Bentz1999} that the diffusion coefficient increases with the size of the cement particles used in the cement paste in the high-porosity region while remaining largely independent of the cement particle size in the low-porosity region.

In this model, we demonstrated that the assumed size of the colloidal CSH impacts the diffusion coefficient and influences the higher-order moments of the displacement probability distribution, such as the shape of the distribution function and the non-Gaussian parameter.
Besides the size effect of the colloidal CSH, an examination of the influence of the assumed CSH shape may also be significant.
Zhang et al. \cite{Zhang2022} recently revealed that the effect of the shape of cement particles (elliptical or spherical) on the chloride diffusion coefficient is limited.
However, it is feasible that the shape of the cement particles may affect the shape of the probabilistic distribution of diffusion displacement, even in the system studied by Zhang et al. \cite{Zhang2022}.
If we adopt Jennings's CM-II model for CSH morphology in modeling, the CSH should possess the shape of an ellipsoid rather than perfectly spherical.
This discussion warrants further investigation.
Hence, it can be inferred that some microstructure elements impact the shape of the probabilistic distribution of diffusional displacement, despite their limited influence on the diffusion coefficient.
Because the tail of the diffusion distribution significantly influences the reliability (durability) assessment of reinforced concrete structures, it is vital to consider not only the diffusion coefficient but also the shape of the displacement distribution in any theoretical, numerical, or empirical approach.

The analysis results in this study indicate that \ce{O_2} diffusion in cement pastes exhibits non-Gaussian behavior for very short timescales, as shown in Figure \ref{fig:ngp_o2_cs}.
However, the duration of this non-Gaussianity could be longer, depending on the heterogeneity of the diffusion medium.
It is important to note that in the context of cement and concrete, the characteristic length referred to in this study should be understood as the cubic root of the REV of the target medium being analyzed.
The presence of interfacial transition zones or the existence of larger aggregates or macro-scale construction defects can undoubtedly increase the characteristic length of the system, thereby extending the persistent timescale of the non-Gaussian behavior that should impact the results of the long-term reliability analysis of concrete structures.

Liu, Liu, and Zhang \cite{Liu2020CCR} conducted a study investigating the dynamics of \ce{CO_2}, \ce{O_2}, and \ce{H_2} in dry cement paste using the lattice Boltzmann method. In their study, a heterogeneous virtual structure of cement paste was constructed using well-developed hydration codes. They determined the diffusion coefficient as a function of porosity; however, the probability density of the displacement $G(x; t)$ or the non-Gaussian parameter $\alpha$ was not examined. The theoretical model presented here is akin to their system; thus, their system exhibits a similar non-Gaussianity of $G(x; t)$ and a non-negligible non-Gaussian parameter. In other words, their simulation methodology can be used to verify our theoretical results.

The current study addresses \ce{O_2} diffusion as a case in point; however, because oxygen is not highly soluble in water, the estimates can be readily extrapolated, even if the cement paste is in a humid environment (i.e., a non-negligible amount of free or physically adsorbed water in capillary pores).
The correction could be accomplished by incorporating the relationship between relative humidity and water adsorption layer thickness \cite{Bentz1995} into estimating the diffusion coefficient of Knudsen diffusion. Additionally, the dynamic change in the specific surface area of CSH in response to changes in relative humidity may affect the slow diffusion coefficient $D_s$, which may need to be considered in some cases.
The extrapolation of the current calculations for the diffusion of other gases, such as \ce{H_2} and Xe, which have low solubility in water, can be easily accomplished by considering the size and mass of the gas molecules like that described by Sercombe et al. \cite{Sercombe2007}.
However, slightly altering the model may be necessary to extend the 2SFD model to \ce{CO_2} diffusion (another crucial aggressive gas species).
Because of its high solubility in water, \ce{CO_2} must be addressed as a diffusion phenomenon in conjunction with the local solubility equilibrium, or it may be immobilized through an \textit{in-situ} carbonation reaction that occurs in the pore solution or inside the CSH gel.
To take this into account, it is essential to incorporate as a third state a stochastic process that transitions on a time scale so long in comparison to the observation time in which the diffusion coefficient is virtually zero. However, the trapped time can be considered effectively infinite.
The transition rate to the additional state should be linked to the carbonation reaction rate; thus, incorporating the Papadakis model and other relevant models could be beneficial for analyzing \ce{CO_2} diffusion in the framework of the multi-state fluctuating diffusivity model.
In addition, a comparable methodology is necessary when addressing the diffusion problem of chloride ions, as it also necessitates the consideration of the effects of chloride binding.
Even when it is expanded to a three-state (even for the extension of a multi-state model), as long as the eigenvalues and eigenvectors of matrix Q in Eq. ~\eqref{eq:isf_general} (the 2SFD model in Eq. ~\eqref{matrixq}) is obtained, all the other calculations can always be performed.
The simplicity of the mathematical structure is another benefit of this model.
The application to the diffusion phenomenon of chemical species (\ce{CO_2}, \ce{Cl-}), which is more critical and reactive for the durability of concrete structures, will be discussed in a future publication as ongoing research in the near future.

\section{Conclusion}
This work applies the analytical method of fluctuating diffusivity to study gas diffusion in cementitious materials.
Note that fluctuating diffusivity is not in opposition to the time-dependent diffusivity approach, reflecting the long-term effects of changing diffusion media, such as prolonged hydration reactions, pore closure due to carbonation, and cracking. Instead, the target timescale is significantly different between the two approaches.
The fluctuating diffusivity framework effectively analyzes the diffusion of small molecules in cementitious materials. The diffusivity may fluctuate spatiotemporally owing to the heterogeneous nature of the diffusion medium and has potential applicability to various diffusion phenomena in these materials.
Our theoretical results of the 2SFD model provide a reasonable description of the diffusion coefficient of \ce{O_2} in colloidal CSH, as measured in previous studies, by estimating the input parameters from the variables in the target systems.
Furthermore, the 2SFD model highlights the presence of non-Gaussian diffusion, which can be attributed to the heterogeneous microstructure of cement pastes.
The persistent timescale of non-Gaussianity essentially depends on the size of the REV of the target media being analyzed (it relies on the presence of interfacial transition zones, the existence of aggregates, etc.).
The presence of non-Gaussianity in the displacement distribution, characterized by heavier tails than those of the Gaussian distribution, is critical for accurately evaluating the long-term reliability probability of reinforced concrete structures. The deviation in the shape of the tail of the Gaussian distribution obtained when solving the diffusion equation using a comparable diffusion coefficient in the 2SFD model may lead to an underestimation of the reliability of the conventional method.
In addition, while some numerical approaches utilize the lattice Boltzmann methods or random walk methods on virtual microstructures generated by previously established hydration models, it is important to acknowledge that there is still ample scope for improvement.
In this regard, the development of a more conceptual stochastic model, such as the 2SFD model rooted in statistical physics, for the examination of diffusion phenomena in cementitious materials from a micro perspective, which can be solved analytically owing to its straightforward theoretical framework, in addition to the structure-based model, would be of great significance to the field of cement and concrete materials research.
This work provides novel insights into the diffusion of small molecules in cement and concrete materials and has the potential for further applications in cement and concrete research.

\section*{Acknowledgement}
F.N. was supported by the Grant-in-Aid for JSPS (Japan Society for the Promotion of Science) Fellows (Grant No. JP21J21725). T.I. was supported by Grant-in-Aid for JSPS (Japan Society for the Promotion of Science) Fellows (Grant No. JP22J00300).

\bibliographystyle{elsarticle-num}
\bibliography{reference}

\end{document}